\documentclass{article}

\usepackage{microtype}
\usepackage{graphicx}
\usepackage{subfigure}
\usepackage{booktabs} 
\usepackage{amsmath}
\usepackage{amsfonts}       

\usepackage{hyperref}

\newcommand{\ignore}[1]{}

\usepackage[accepted]{mlsys2020}

\mlsystitlerunning{IMPALA: Low-Latency, Communication-Efficient Private Deep Learning Inference}

\begin{document}

\twocolumn[
\mlsystitle{IMPALA: Low-Latency, Communication-Efficient \\ Private Deep Learning Inference}




\begin{mlsysauthorlist}
\mlsysauthor{Woo-Seok Choi}{snu}
\mlsysauthor{Brandon Reagen}{nyu}
\mlsysauthor{Gu-Yeon Wei}{har}
\mlsysauthor{David Brooks}{har}
\end{mlsysauthorlist}

\mlsysaffiliation{snu}{Seoul National University, Seoul, South Korea}
\mlsysaffiliation{nyu}{New York University, NY, USA}
\mlsysaffiliation{har}{Harvard University, MA, USA}


\mlsyskeywords{Machine Learning, MLSys}

\vskip 0.3in

\begin{abstract}\label{sec:abstract}

This paper proposes Impala, a new cryptographic protocol 
for private inference in the client-cloud setting.
Impala builds upon recent solutions that combine the complementary strengths of 
homomorphic encryption (HE) and secure multi-party computation (MPC).
A series of protocol optimizations are developed
to reduce both communication and performance bottlenecks.
First, we remove MPC's overwhelmingly high communication cost
from the client by introducing a proxy server and developing a low-overhead key switching technique.
Key switching reduces the clients bandwidth by multiple orders of magnitude,
however the communication between the proxy and cloud is still excessive.
Second, to we develop an optimized garbled circuit 
that leverages truncated secret shares for faster evaluation and less proxy-cloud communication.
Finally, we propose sparse HE convolution to reduce the computational bottleneck of using HE.
Compared to the state-of-the-art, these optimizations provide a
bandwidth savings of over $3\times$ and speedup of $4\times$ 
for private deep learning inference.

\end{abstract}
]



\printAffiliationsAndNotice{}  

\section{Introduction}
\label{sec:intro}

\begin{figure*}[t]
\begin{center}
\centerline{\includegraphics[width=\textwidth]{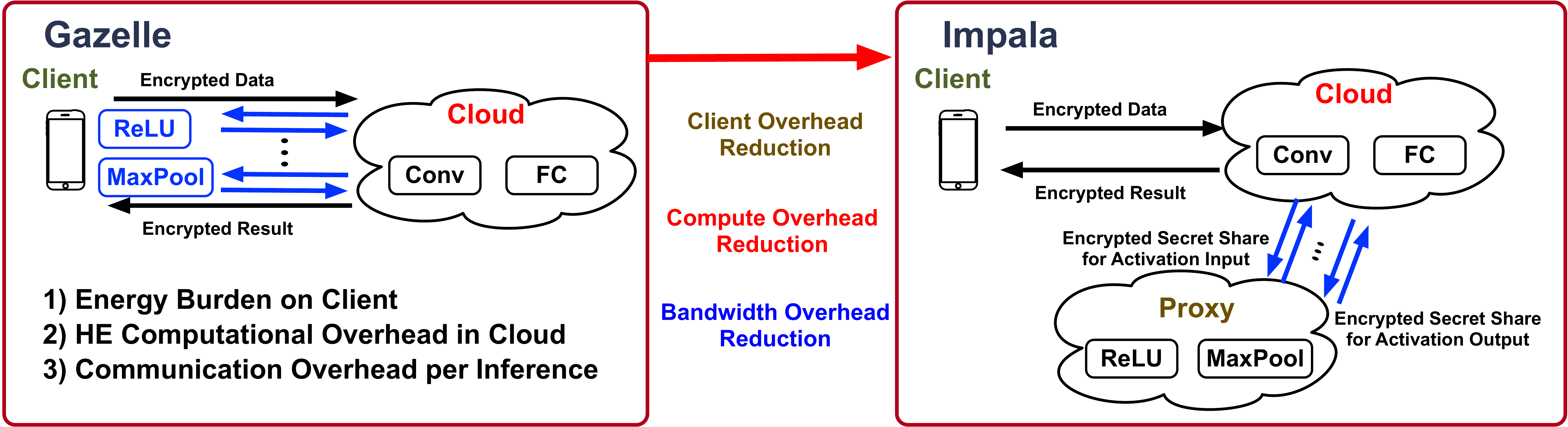}}
\vskip -0.1in
\caption{Overview of Impala and contributions.}
\label{fig:proposed}
\end{center}
\vspace{-2em}
\end{figure*}

Countless services and applications have benefited from deep learning.
The utility of a model is often improved when more
information is shared, both to train better models and to provide meaningful predictions.
However, recent concerns over users' data use threaten today's deep learning workflows and
have galvanized efforts to improve privacy in deep learning.
This has been especially visible in the client-cloud compute setting,
where clients' data are sent to backend servers for processing,
including Machine-Learning-as-a-Service (MLaaS) applications.

There are two general ways to achieve private inference.
One approach is to leverage system-based solutions, 
e.g., trusted execution environments (TEEs) or client-local inference.
System-based solutions are highly attractive as they are practically deployable today.
While efficient, the drawback of these solutions is a relaxed threat model. 
For example, local execution protects client inputs but not the cloud's models,
and TEEs have been shown vulnerable to side-channel attacks~\cite{costan2016intel}.
The second approach is to leverage cryptocraphic techniques to devise protocols
then enable computation directly on encrypted data.
Techniques including secure multi-party computation (MPC) and homomorphic encryption (HE)
protect against stronger threat models than system-based approaches at the expense
of significantly higher communication and computation overheads.

The goal of this paper is to enable cryptographically-secure private inference 
without excessively burdening the client.
We observe that while existing work has made significant efforts 
in developing secure protocols for cryptographic inference
they all introduce significant cost, both computation and communication, for the client.
For example, recent private inference protocols based on HE
have shown that significant cloud performance benefits are possible
with software and hardware optimizations~\cite{juvekar2018gazelle, reagen2020cheetah}.
However, the solutions neglect to resolve the communication requirements
of the client, which in the case of CIFAR-10 exceed 1\,GB \emph{per inference}!
Our philosophy is that because the server/service provider are profiting from the clients'
data, they should handle the increased processing load of private execution, not the client.

Existing cryptographic protocols for private inference take a hybrid approach
to combine the complementary strengths of both MPC and HE techniques for different computations.
Here we review solutions for both MPC+HE and MPC+MPC techniques to highlight
that, while this paper proposes an MPC+HE protocol, the problem is ubiquitous.
In MPC+HE protocols, Garbled Circuits (GCs) are used to process non-linear 
(i.e., ReLU and MaxPooling) inference
operations and HE linear (i.e., FC and Conv).
The benefit of combining HE and MPC is that the client needs to do little 
additional processing for secure inference as HE is performed by the client.
Because HE is used, a high performance overhead is incurred by the cloud,
also the GC requires significant communication overhead.
In MPC+MPC solutions, a second type of MPC, called secret sharing, is
used for linear layers in place of HE.
The benefit of secret sharing is that the cloud side computational overheads are
significantly less than with HE.
However, it does require the client to also compute the same computation
and incur the high communication overheads of generating multiplicative secrets and the GC.
This paper focuses on MPC+HE as it introduces the least overhead to the client.

While Gazelle is a significant first-step and shows the best performance published so far,
Gazelle's implementation suffers from three addressable problems (Fig.~\ref{fig:proposed}): 
1) The client incurs large communication burden as due to the MPC
protocol invoking the user's device after each linear layer;
2) High computational overhead remains for HE;
3) Each inference requires large amounts of communication 
(e.g., \textbf{1.2\,GBs} using CIFAR-10).

This paper proposes Impala, a cryptographically-secure private inference protocol, 
which improves upon each of the three aforementioned shortcomings
using a series of co-designed optimizations (see Fig.~\ref{fig:proposed}).
First, we optimize the secure inference protocol by introducing a proxy server and enabling interactions between
the proxy and cloud directly with \textit{key swtiching}.
This protocol all but eliminates the inhibiting client-side communication requirements of Gazelle, 
saving multiple orders of magnitude of client-cloud communication.
Next, on top of the newly proposed HE-based linear kernel implementation algorithm in Cheetah~\cite{reagen2020cheetah},
we leverage the weight sparsity in convolutional neural networks (CNNs) to improve the latency of processing linear layers using HE. 
In order to reduce the communication bandwidth required for evaluating non-linear functions securely in the proposed protocol, 
the GC size is reduced by truncating the secret shares of cloud and proxy.
Both the HE and GC optimizations improve computational efficiency without compromising model accuracy, 
i.e., inference accuracy of a fixed-point CNN in plaintext is maintained in Impala.

This paper makes the following contributions:
\begin{itemize}
    \item 
    Impala, an optimized protocol for 
    private inference in the client-cloud setting is proposed.
    The protocol introduces a semi-honest proxy to efficiently evaluate non-linear functions with MPC and reset HE noise from FC/CONV layers run on the cloud.
    The main idea is to adapt a key switching based re-encryption technique with minimal computational overhead to
    enable the proxy and cloud to work together directly and securely, 
    i.e, without the client's secret key. 
    Compared to state-of-the-art, this reduces the client's bandwidth by more than 3500$\times$.

    \item 
    To further reduce the computation and communication overhead of MPC,
    we propose an optimized implementation of non-linear inference operators (e.g., ReLU and MaxPool) in garbled circuits with truncated secret shares.
    This technique reduces the total communication bandwidth by over $3\times$ compared to state-of-the-art.
    In addition, it enables Impala to support fixed-point arithmetic used for CNN inference.
    
    \item 
    We carefully examine two HE-based linear kernel implementations proposed by Gazelle and Cheetah,
    and analyzed the total HE noise after linear layer computation.
    We find out that Cheetah's implementation outperforms Gazelle in the client-cloud setting especially for deep CNNs.
    In order to further reduce the latency of processing CNNs using HE, weight sparsity in CNNs is leveraged.
    Smartly mapping CNN operators onto HE primitives enables computations be reduced in direct proportion to kernel sparsity.
    This sparsity optimization provides 4$\times$ inference speedup over state-of-the-art.
    
\end{itemize}

\nocite{appleLDP,ding2017collecting,8416855,tramer2018slalom,bos2014private}
\section{Assumptions and Related Work}\label{sec:assumption}

\subsection{System Overview}\label{sec:system_overview}
In this paper, we assume that a client-cloud compute setting where
a user is running an application on a client device (e.g., a smartphone)
and the application relies on the cloud for high-performance computing.
Specifically, when applications use deep learning, we assume that inference is off-loaded to the cloud:
the client generates new data, sends it to the cloud, the cloud performs the inference,
and returns the result to the client.
This setting further assumes that the client and cloud want to keep their data/model private/protected.
The client would prefer their data is kept private and, likewise, the cloud its trained model.
The client-cloud setting is similar to common ML or Prediction as a Service (ML/PaaS) model---our privacy-preserving protocol holds for these scenarios as well.

When privacy is a concern, we assume that:
the client encrypts its data, this data can only be decrypted by the client, and the results are returned encrypted;
the cloud/application provider's models are kept private.
To be more specific, the weights of trained models are not leaked.

\subsection{Threat Model}\label{sec:threat_model}

Two adversary models are typically considered.
In the {\em Semi-honest}, or {\em honest-but-curious}, model,
corrupt parties try to learn other parties' private information while following the protocol honestly.
A {\em malicious} adversary is a corrupt agent that deviates from the protocol in arbitrary ways.
We assume the \textit{honest-but-curious} threat model as it is
commonly used in evaluating privacy-preserving ML protocols.
Moreover, our assumed threat model is the same as prior works 
(i.e., honest-but-curious and honest majority assumption, detailed below) on privacy-preserving inference
\cite{rouhani2017deepsecure, liu2017oblivious, riazi2018chameleon, juvekar2018gazelle}.

Impala is specifically designed for three parties (3PC) (client, cloud, and proxy),
and we assume that at most one party can be corrupt
(common in 3PC~\cite{bogdanov2008sharemind, bogdanov2012high, araki2016high}).
That is, we consider the honest majority case
where two parties do not collude to disclose the other party's private information. 
In our protocol both the {\em client's data} and 
{\em cloud's model parameters} are kept private under this threat model.
We note that the proxy (introduced as a protocol optimization, see Section~\ref{sec:proposed})
is not required to be a trusted party, and the privacy of the data is preserved
as long as the proxy and the cloud do not collude as neither knows the client's secret key.
I.e., Impala ensures that neither the cloud nor the
proxy alone can learn any information about client's private data. 

The security proof is sketched as follows. 
The cloud observes only ciphertexts encrypting client's data, and processes it using homomorphic encryption.
The proxy observes ciphertexts from the \textit{cloud} and additive secret shares that look random 
(under honest majority assumption) after decryption.
Since both ciphertexts and secret shares provide indistinguishability under chosen plaintext attack (IND-CPA), 
the cloud/proxy cannot learn anything from what they observe,
and the privacy of the client's data and the cloud's model are guaranteed.
Impala provides the same level of security as other cryptographic solutions~\cite{rouhani2017deepsecure,liu2017oblivious,riazi2018chameleon,juvekar2018gazelle}
while improving the performance significantly.

\subsection{Related Work}\label{sec:related_work}

The growing interest for privacy in machine learning
has fueled many prior publications on cryptographically-secure private inference.
Solutions can be categorized into two groups: HE only~\cite{gilad2016cryptonets,hesamifard2017cryptodl,brutzkus2019low},
or MPC-based~\cite{rouhani2017deepsecure,liu2017oblivious,riazi2018chameleon,juvekar2018gazelle}.
Each has advanced the field with significant performance improvements.
However, they all suffer from either accuracy loss due to nonlinear activation function approximation
or high communication and computational complexities that scale with network depth.

HE suffers from evaluating the functions that are hard to approximate using multiplication and addition, 
e.g. nonlinear activations (ReLU, MaxPool) necessary for highly-accurate deep learning models.
CryptoNets~\cite{gilad2016cryptonets}, CrytoDL~\cite{hesamifard2017cryptodl}, and LoLa~\cite{brutzkus2019low} mitigated this problem by replacing ReLU with low-order polynomials that can be practically computed using HE.
However, even with square activations, these solutions require inordinately large
HE parameter (e.g., ciphertext modulus $q$ of around 1000 bits) to allocate a noise budget capable of processing an inference.
This result in extreme run times and the technique is only applied to small models.
In addition, using square activation requires re-training the entire model.

Given the compute complexities of HE, some works have proposed MPC-based schemes as an alternative~\cite{orlandi2007oblivious,7958569,rouhani2017deepsecure,liu2017oblivious,riazi2018chameleon,juvekar2018gazelle}.
In MPC, the involved parties, e.g., the client and cloud, evaluate inferences interactively.
In \cite{orlandi2007oblivious,7958569} linear operations use a Paillier cryptosystem~\cite{paillier1999public},
which is additively homomorphic.
Others have proposed leveraging secret sharing for linear operation and generating secrets offline to improve performance~\cite{liu2017oblivious}.

Recently, low-latency 3PC protocols such as \cite{mohassel2018aby,araki2016high} were introduced to improve the performance over 2PC, 
but, because 3 parties should be involved in computation, using them in the client-cloud setting requires overall 4 parties (e.g., client, cloud, proxy 1, and proxy2).
On the other hand, in Impala only two parties (cloud and proxy) are required to compute,
which is the main difference from general 3PC frameworks.
In addition, other 3PC protocols in the client-cloud setting guarantee privacy when none of the 4 parties colludes with someone else.
Compared to our threat model, this clearly demands stronger assumption. 
Moreover, the inference latency of Impala is comparable to the performance of ABY3~\cite{mohassel2018aby}.
Note that the performance improvement in Impala is not due to introducing the proxy in the protocol but due to sparse HE convolution and optimized GC implementation.
The proxy's job is to perform computation on behalf of the client, thereby reducing the energy burden on the client device.
Without the proposed techniques on HE convolution and GC,
the overall inference latency actually increases due to the additional key switching procedure required for securely adding the proxy in the protocol.

Most relevant to our work is Gazelle~\cite{juvekar2018gazelle},
which proposed using SIMD (single instruction, multiple data)\footnote{SIMD is also called {\em batch encoding} in homomorphic encryption literature.} 
HE for linear layers and Yao's GC~\cite{yao1986generate} for the nonlinear activation functions.
The communication and computational complexity of the GC method are roughly proportional to the number of Boolean logic gates required to implement the activation functions.
Some activation functions such as ReLU or MaxPool can be expressed using boolean circuit efficiently,
which results in better performance compared to HE-based evaluation.
We significantly improve the performance by reducing the amount of communication and computation using the proposed techniques.
Moreover, we aim to avoid the burden on the client incurred by 2PC by adding a proxy securely in Impala.

\section{Background}\label{sec:background}

This section introduces BFV homomorphic encryption and garbled circuits,
both core technologies for Impala.
For more details, see~\cite{brakerski2012fully,fan2012somewhat,rouhani2017deepsecure}.

\subsection{Homomorphic Encryption (HE)}
\label{sec:background_he}
HE is a cryptographic approach for preserving privacy that enables direct computation of arbitrary functions on encrypted data.
Known HE schemes support addition and multiplication, and
the support of arbitrary functions comes from the fact that
any arithmetic function can be decomposed into additions and multiplications (e.g., Taylor-series expansion).
Since Gentry first demonstrated fully homomorphic encryption (FHE)~\cite{Gentry:2009:FHE:1536414.1536440}, many have proposed ways to
improve performance~\cite{Gentry:2010:CAF:1666420.1666444,brakerski2014leveled,gentry2012fully,gentry2013homomorphic,bos2013improved,brakerski2012fully,fan2012somewhat}. 

In general, HE works by encrypting data with noise for security,
and noise increases with each arithmetic operation performed on the encrypted data.
Correct computation fails if this noise increases beyond a certain threshold (aka noise budget),
which is decided by the encryption parameters.
This means, for a set of parameters,
HE can only evaluate arithmetic circuits with bounded depth (called leveled-HE or LHE).
While {\em Bootstrapping} is an intermediate noise-resetting technique, it is extremely computationally expensive and most HE schemes currently being used are LHE~\cite{wu2012using,bos2014private,gilad2016cryptonets,hesamifard2017cryptodl}.

\subsubsection{BFV Scheme and Encryption Parameters}

In BFV, the plaintext space is a ring,
$R_t = \mathbb{Z}_t[x]/(x^n+1)$, where the elements are polynomials of degree less than $n$
(typically chosen as a power of 2) 
with integer coefficients in the range $(-\frac{t}{2}, \frac{t}{2}]$ for some plaintext modulus $t > 1$.
Given $q$ (ciphertext modulus) to be another integer larger than $t$,
a plaintext polynomial is encrypted to a ciphertext 
composed of two polynomials in $R_q = \mathbb{Z}_q[x]/(x^n+1)$.
As stated earlier, randomly sampled noise following bounded discrete Gaussian distribution is added during encryption,
and this noise in the ciphertext keeps increasing as more operations are applied to the ciphertext. 
After processing, decryption returns a correct output if the largest magnitude of coefficients of noise polynomial $\mathbf{v}$ is less than $q/2t$
(i.e. $||\mathbf{v}||_{\infty} < q/2t$).
Since $||\mathbf{v}||_{\infty}$ is critical for correct decryption, it is important to understand how each arithmetic operation on a ciphertext affects $||\mathbf{v}||_{\infty}$ (see Section \ref{sec:bg_hebfv}).
In a fresh ciphertext (i.e., right after encryption), 
$||\mathbf{v}||_{\infty} \approx 12n\sigma$ ($\sigma$: noise standard deviation of Gaussian distribution), 
so smaller $\sigma$ provides a larger noise budget, but it also makes the encryption less secure.
Prior research explains how to choose $\sigma$ for a given security parameter and $(n, q)$ \cite{albrecht2015concrete}.

Properly choosing encryption parameters ($n, t, q)$ can improve performance.
Instead of encrypting a single data (scalar) into a single ciphertext, 
it is possible to pack $n$ data into one plaintext,
thus one ciphertext, and to process the packed data in a SIMD fashion; 
view each ciphertext to have $n$ data slots, similar to an $n$-dimensional vector.
This allows to process $n$ packed data in parallel with a single operation on a ciphertext, 
so the HE overhead of BFV are amortized, thereby improving the performance by a factor of $n$.
In addition, this reduces the ciphertext size, saving communication between the client and cloud.
To enable SIMD, $t$ needs to be prime with $t \equiv 1 \mod 2n$.
Under this condition, by the Chinese Remainder Theorem (CRT), every $n$-dimensional integer vector in $\mathbb{Z}_t^n$ can be one-to-one mapped to a unique polynomial plaintext in $R_t$,
i.e., $\mathbb{Z}_t^n$ is isomorphic to $R_t$.
For more details on SIMD, see \cite{smart2014fully}.
In addition, by choosing $q$ to be prime with $q \equiv 1 \mod 2n$, 
multiplication between polynomials in $R_q$ can be accelerated using the Number Theoretic Transform (NTT).
This reduces the complexity of polynomial multiplication from $O(n^2)$ to $O(n\log n)$.
Impala chooses $(n, t, q)$ to take advantage of both SIMD and NTT.

\subsubsection{Basic Homomorphic Operations in BFV}
\label{sec:bg_hebfv}


\textbf{Addition:} Two ciphertexts $\mathsf{ct^0, ct^1}$ (encrypting $\mathbf{m_0, m_1}$) can be added homomorphically as follows: ($\mathsf{ct^i} = (\mathbf{ct_0^i, ct_1^i})$ with $\mathbf{ct_0^i, ct_1^i} \in R_q$)
\begin{equation}
    \mathsf{ct_{add}} = \mathsf{ct^0 + ct^1} = \mathbf{(ct_0^0 + ct_0^1,\; ct_1^0+ct_1^1)}.  \label{eq:simdadd}
\end{equation}
It can be verified that decrypting $\mathsf{ct_{add}}$ results in the addition of $\mathbf{m_0}$ and $\mathbf{m_1}$, while the noise grows additively, 
i.e., $\mathbf{v_{add}} \approx \mathbf{v^0} + \mathbf{v^1}$.
With SIMD, single ciphertext addition performs parallel $n$ slot-wise data addition.


\textbf{Plaintext Multiplication:} A plaintext $\mathbf{m_w} \in R_t$ containing model weights can be multiplied to a ciphertext $\mathsf{ct}$ as:
\begin{equation}
    \mathsf{ct_{mult}} = \mathbf{(m_wct_0,\; m_wct_1)}.    
    \label{eq:simdmult}
\end{equation}
It can be verified that decrypting $\mathsf{ct_{mult}}$ results in the multiplication of $\mathbf{m}$ and $\mathbf{m_w}$ (slot-wise multiplication with SIMD), while the noise grows multiplicatively, 
i.e., $\mathbf{v_{mult}} \approx \mathbf{m_w v}$.
Since $||\mathbf{m_w}||_{\infty}$ can be as large as $t/2$, after each plaintext multiplication, noise can grow by a factor of $nt/2$ in the worst case.
Note that even though the packed scalars are small
(e.g., a weights vector is composed of 0's and 1's in the extreme case),
when they are converted to a plaintext $\mathbf{m_w}$ using CRT,
$||\mathbf{m_w}||_{\infty}$ can be close to $t/2$ and cause significant increase in noise.


\textbf{Automorphism:} When a ciphertext contains multiple data for SIMD operation, slot rotation is required to allow computation between data in different slots.
For example, if we want to add data at different slots in a ciphertext,
we can add them after rotating one data to the common slot.
Automorphism rotates slots in a cyclic manner. 
The detailed procedure can be found in \cite{brakerski2014leveled,wu2012using}.
One distinct feature is that the client should provide the public (evaluation) keys to the cloud to perform automorphism,
and each key corresponds to a specific rotation.
Automorphism increases noise additively, 
but the amount of noise is decided by a parameter $w_A$ (ciphertext coefficient decomposition base) introduced by automorphism.
Choosing smaller $w_A$ adds less noise but results in more number of decomposed polynomials, thereby increasing the latency.

\subsection{Garbled Circuit (GC)}
\label{sec:background_gc}

Yao's GC~\cite{yao1986generate} is a cryptographic protocol 
that allows two parties to jointly compute a function over their private data
without learning the other party's data.
For GC, the computed function should be represented as a Boolean circuit 
with logical gates (e.g., XOR, AND, etc.), and 
the inputs from each party are represented as input wires to the circuit. 
Then, the protocol proceeds as follows: 
1) one of the parties (Garbler) garbles the circuit and generates the garbled table. 
2) The garbled table is transferred to the other party (Evaluator) along with the randomized labels 
corresponding to both parties' input using {\em Oblivious Transfer} (OT)~\cite{ishai2003extending}.
3) Evaluator evaluates the table to get the result.

Since the amount of required computation and communication is proportional to the Boolean circuit size,
GC is able to efficiently process functions that can be represented as Boolean circuits compactly compared to HE.
Especially, widely-used activation functions in CNNs like ReLU or MaxPool are hard to approximate using addition or multiplication
but they can be efficiently expressed using Boolean circuits.
Therefore, most of the MPC-based works~\cite{rouhani2017deepsecure,liu2017oblivious,riazi2018chameleon,juvekar2018gazelle} used GC to evaluate ReLU and MaxPool.
Moreover, recently GC has benefited from significant performance improvements~\cite{zahur2015two,ishai2003extending,bellare2013efficient}, 
making it suitable for evaluating the activation functions of neural networks.
\section{Proposed Solution}
\label{sec:proposed}

In this section we present the techniques---{\em sparse homomorphic convolution}, {\em re-encryption with key switching}, and privacy-preserving activation evaluation using {\em GC with truncated secret shares}---
to improve computational and communication efficiency tailored for CNN inference.
Impala, a novel privacy-preserving protocol for deep learning inference using the described techniques, is then proposed.

\begin{figure*}[t]
\begin{center}
\centerline{\includegraphics[width=\textwidth]{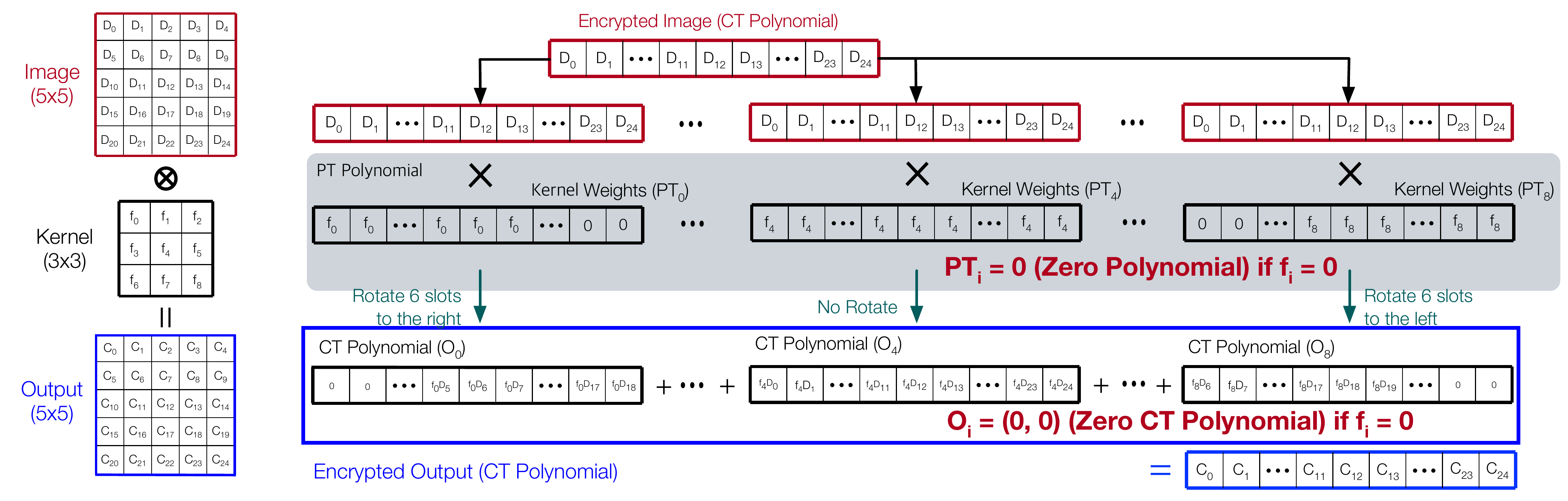}}
\caption{Homomorphic convolution with a sparse kernel.}
\label{fig:conv}
\end{center}
\vspace{-1em}
\end{figure*}

\subsection{Sparse Homomorphic Convolution}
\label{SC}

We first describe how homomorphic convolution is implemented 
and show that a trained model with sparse kernels reduces the computational complexity,
which is proportional to the sparsity level.

\begin{table}[t]
\caption{Output HE noise after CONV layer.}
\begin{center}
\begin{small}
\begin{tabular}{cc}
\toprule
 & Output Sub-Gaussian Noise Parameter \\
\midrule
Gazelle &  $f_w^2\sqrt{c_i}\sqrt{2+\frac{w_{A,G}^2l_{A,G}}{4}}\frac{t}{2}n\sigma$  \\
Cheetah &  $f_w^2\sqrt{c_i}\sqrt{2 + \frac{w_{A,I}^2l_{A,I}}{t^2n}}\frac{t}{2}n\sigma$  \\
\bottomrule
\end{tabular}
\end{small}
\end{center}
\label{t:noise_conv}
\vspace{-1em}
\end{table}

\subsubsection{Convolution Implementation Comparison}

The same approach as Cheetah~\cite{reagen2020cheetah} was used to implement the homomorphic convolution
because Gazelle's implementation suffers from large computational overhead to process large CNN models in the client-cloud compute setting.
(Homomorphic convolution can be also performed following LoLa~\cite{brutzkus2019low},
but it must leak the information of the kernel size to the other party 
and also increases the ciphertext size by a factor of the kernel size.)
The main drawbacks of Gazelle are that 
1) latency increases substantially when the matrix or kernel size for FC and CONV layers become large, and
2) the {\em hoisting} optimization technique~\cite{halevi2018faster} cannot be used without huge energy burden on the client during the key generation phase.
Gazelle exploited hoisting to rotate the same input ciphertext and reduce the number of NTT for automorphism through which it achieves large speed-up.
However, this can be applied only when the desired rotation can be performed with single automorphism,
and, as explained in Section~\ref{sec:bg_hebfv}, all the evaluation keys corresponding to the desired rotation should be provided by the client.
In other words, to get the benefits of hoisting, the client needs to generate and transmit $(n-1)$ evaluation keys,
requiring large amount of client-side computation and communication,
and this is not desirable in the client-cloud model.
On the other hand, since Cheetah does not take advantage of hoisting, the client does not have to generate large number of keys.

The homomorphic convolution in Cheetah is performed as follows.
For instance, as shown in Fig.~\ref{fig:conv}, 
if the kernel size is 9 then each element in the output channel is the sum of 9 scaled input elements.
One of the nine can be obtained by multiplying $f_4$ commonly to all the slots,
while the remaining eight are added using plaintext multiplication \textit{and} automorphism.
Thus, a homomorphic convolution with kernel size of 9 requires a total of:
9 plaintext multiplication, 8 automorphism, and 8 addition.
The zeros found in plaintext slots (e.g. PT0 in Fig.~\ref{fig:conv}) are a result of automorphism working in a cyclic manner.
For example, after $f_0$ is multiplied to $D_0$, 
we need to rotate the data by 6 steps to the right to add the result in the slot 6 ($C_6$ in the output).
While doing so, however, $D_{19}$ comes to the slot 0 but $D_{19}$ should not be added to the slot 0.
To prevent this, zeros are selectively added in the plaintext slots. 
Note that, in Gazelle, automorphism is performed before plaintext multiplication to exploit hoisting.
However, because the sequence of the HE operations is different, the output noise becomes different, 
resulting in different amount of computation required for convolution.

In order to make a fair comparison between two implementations,
we first represent the distribution of the output noise as a function of encryption ($n, q, t, \sigma, w_A)$ and CONV layer parameters $(c_i, c_o, f_w, w)$,
where $(c_i, c_o, f_w, w)$ represent the number of input and output channels, width of the kernels and images, respectively.
To this end, we note that the noise in the fresh ciphertext has sub-gaussian distribution with the parameter of $\sqrt{2n}\sigma$,
and plaintext multiplication amplifies the sub-gaussian parameter by a factor of $\sqrt{n}t/2$.
Noise added by automorphism also follows sub-gaussian with the parameter of $\sqrt{l_An}\sigma w_A/2$.
Using these, the output noise parameter can be expressed as in Table~\ref{t:noise_conv}.
We can see that under the same security setting (i.e., same $(n, t, q, \sigma)$), 
the decomposition base for automorphism $w_A$ in Cheetah can be chosen around $\sqrt{n}t\times$ larger than that of Gazelle,
which saves huge amount of computation in large CNNs.
The speed-up becomes even more substantial in the client-cloud model, where the client provides limited amount of evaluation keys and hosting cannot be exploited (see Table~\ref{t:com_fc} and \ref{t:com_conv}).
Therefore, Cheetah's implementation is adopted in Impala.

\subsubsection{Leveraging Sparsity}

To further improve the performance, model weight sparsity is leveraged in Impala.
Note that, although graphically represented in Fig.~\ref{fig:conv},
each PT's and CT are actually high-degree polynomials,
and all the operations are performed by a sequence of multiplication and addition over the polynomial ring described in Section~\ref{sec:background_he}.
The computational complexity 
of homomorphic convolution with the kernel parameters $(c_i, c_o, f_w, w)$ is $O(c_ic_of_w^2)$. 
It can be easily seen from Fig.~\ref{fig:conv} that if an element in the kernel is zero,
then automorphism, plaintext multiplication, and addition corresponding to the zero element can be skipped
because the plaintext corresponding to zero is also mapped to zero polynomial in the plaintext space.
In other words, if the sparsity (fraction of zeros) of the entire kernels is $\alpha$ ($0 \leq \alpha \leq 1$),
the amount of computation decreases to $O((1-\alpha)c_ic_of_w^2)$.
Note that, compared to convolution, homomorphic matrix-vector multiplication for FC layers 
can be implemented efficiently following \cite{wu2012using,halevi2014algorithms,juvekar2018gazelle, reagen2020cheetah},
where multiple weights (e.g. rows or diagonals) are packed into a single plaintext.
In this implementation, however, weight sparsity in FC layers does not help reduce the amount of computation.
Since the performance bottleneck is mainly due to homomorphic convolution,
we improve the overall latency by training a CNN with sparse kernels using prior techniques~\cite{zhu2017prune}.

There exists some prior art like Faster CryptoNets~\cite{chou2018faster} that tried to leverage sparsity to make HE faster. 
However, Faster CryptoNets encrypted each pixel into a ciphertext without SIMD (batch encoding) to exploit sparsity,
so even with a highly sparse model, it still suffers from large latency (see Table~\ref{t:mnist}).
Sparsity with SIMD can be leveraged only when zero's are multiplied to all the packed input in a ciphertext, 
so sparsity cannot be used for computing FC layer and it can be used for CONV layer 
only when the homomorphic convolution is implemented as shown in Fig.~\ref{fig:conv}.

\subsection{Re-encryption with Key Switching}
\label{ks}

All prior works employing secure two-party computation
(2PC)~\cite{orlandi2007oblivious,7958569,rouhani2017deepsecure,liu2017oblivious,riazi2018chameleon}
including Gazelle~\cite{juvekar2018gazelle} introduces a large energy burden to the client device that increases proportionally with model depth,
restricting technique applicability to high-performing client devices and limiting the depth of ML models.
To solve this, we introduce an independent and untrusted third-party proxy to the system.
Impala uses the proxy to execute nonlinear layers using GC on behalf of the client.
The challenge is in how the proxy is allowed to process the intermediate result
that it receives from the cloud fully encrypted.
In order to use GC and obtain its private share, the encrypted result should be first decrypted,
which cannot be processed without the client's secret key,
and sharing a key with a third party is unacceptable in practice.

In order to use the proxy without compromising the client's secrete key,
the proxy in Impala uses an independent key ($\mathsf{s_p}$) generated by the client,
and the cloud converts, or re-encrypts, the output ciphertext (encrypting the linear layer results) 
so the proxy can decrypt it using the $\mathsf{s_p}$ only.
To this end, we employ the key switching subroutine found in relinearization or automorphism~\cite{brakerski2014leveled}.
We modify the approach so the proxy can be introduced in the system without compromising the client's privacy.
The procedure is described below for completeness.
For re-encryption, the client provides the cloud with a special public key for key switching as follows: ($w_{SW}$ is a parameter and $l_{SW} = \lfloor \log_{w_{SW}} q \rfloor + 1$)
\begin{align}
    &\mathsf{ReEncKey}:\forall i \in \{0,\dots,l_{SW}-1\}: \mathbf{a_i}\xleftarrow{\$}R_q, \mathbf{e_i}\leftarrow \chi.  \nonumber\\
    &\qquad \qquad 
    \mathsf{pk_{SW}}^{(i)} = ([-(\mathbf{a_i}\mathsf{s_p} + \mathbf{e_i}) + w_{SW}^i\mathsf{s_c}]_q,\; \mathbf{a_i})    
    \label{eq:reenckeygen}
\end{align}
Then, key switching in the cloud is performed as follows:
\begin{align}
    &\mathsf{ReEnc : } \text{ Decompose coeff. of } \mathbf{ct_{1}} \text{ with base } w_{SW}. \nonumber \\
    & \qquad \quad \text{ Output }  \mathsf{ct'} = (\mathbf{ct_{0}',\; ct_{1}'}) \text{ with} \nonumber \\
    & \qquad \qquad
    \mathbf{ct_{0}'} = \mathbf{ct_{0}} + \sum_{i=0}^{l_{SW}-1}(\mathbf{pk_{SW0}}^{(i)}\cdot\mathbf{ct_{1}}^{(i)}), \nonumber \\
    & \qquad 
    \qquad \mathbf{ct_{1}'} = \sum_{i=0}^{l_{SW}-1}(\mathbf{pk_{SW1}}^{(i)}\cdot\mathbf{ct_{1}}^{(i)})
    \label{eq:reenc}
\end{align}
The resulting ciphertext $\mathsf{ct'}$ can be decrypted with only proxy key, and the additive noise incurred by key switching is bounded by $l_{SW}w_{SW}Bn/2$.
For small latency overhead, $w_{SW}$ should be chosen as large as possible.
The idea of adding a proxy securely in protocol design and the concept of proxy re-encryption over various cryptosystems are not new~\cite{ateniese2006improved,10.1145/3128607}. 
The novelty here is that 1) the re-encryption compatible with the BFV scheme is implemented, and 2) more importantly, re-encryption is performed in Impala so that the latency overhead added by key-switching can be minimal.
Note that introducing the proxy in the system increases overall computation (latency) due to this extra key-switching operation.

The key-switching operation must be done 
{\em after} finishing computing a linear layer homomorphically
to minimize the output noise, 
which allows choosing large $w_{SW}$ and minimizing the latency overhead.
If key switching is performed {\em before} linear operators, 
then the noise added by key switching will be amplified with plaintext multiplication, 
leading to a choice of computationally inefficient HE parameters
(e.g. small $w_{SW}$ for small key-switching noise or large $q$ for large noise budget).
Moreover, because key switching needs to be done only once in the first layer, it adds negligible overhead to the overall protocol.
Since all the computation is performed between the cloud and proxy, 
the client's only job is to encrypt private data and decrypt the final result,
making Impala suitable for resource-constrained clients.

\subsection{GC-based Nonlinear Activation Function Evaluation}
\label{gc}

\begin{figure*}[t]
\begin{center}
\centerline{\includegraphics[width=\textwidth]{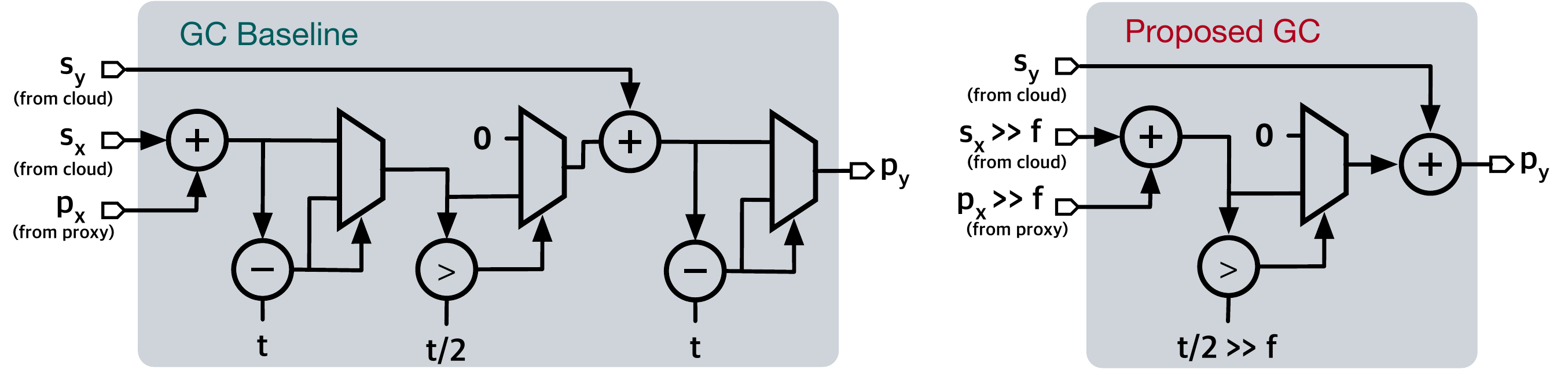}}
\caption{Comparison between the proposed GC for ReLU and baseline implementation~\cite{juvekar2018gazelle}. The proposed implementation supports fixed-point arithmetic and saves communication bandwidth due to smaller circuit size}
\label{fig:opt_gc}
\end{center}
\vspace{-1em}
\end{figure*}
Since the proxy is invoked every time the nonlinear function is evaluated,
Impala {\it does} leak information about the model depth like other MPC-based works~\cite{rouhani2017deepsecure,liu2017oblivious,riazi2018chameleon,juvekar2018gazelle}.
Other parameters (e.g., model weights or kernel size, etc.), however, are kept private.
Thus, Impala guarantees users' data privacy while 
protecting the cloud's models (considered IP today~\cite{zhang2018protecting})
from model-stealing attacks (e.g., \cite{tramer2016stealing}).

To meet this privacy requirement, 
Impala evaluates the nonlinear functions using GC between the cloud and proxy.
The major limitation of GC is its large communication bandwidth,
which is proportional to the circuit size and can become problematic for low-bandwidth networks.
Especially, since the whole process described in Section~\ref{sec:background_gc} should be done for each inference 
(e.g., 6.2\,GB for MiniONN~\cite{liu2017oblivious} and 1.2\,GB for Gazelle to process CIFAR-10),
it is critical to reduce the Boolean circuit size for better performance. 
Another problem in prior works using HE, e.g. CryptoNets, LoLa, and Gazelle, is that fixed-point arithmetic has not been taken into account because rounding (or truncating LSBs) is too expensive to perform in BFV.
To exploit prior works, fixed-point numbers in the model should be mapped to integers,
but without rounding the bit size of each layer output increases exponentially with the model depth.
There exists a HE scheme that supports fixed-point arithmetic~\cite{cheon2017homomorphic},
but we propose an efficient technique that can enable fixed-point arithmetic and lower GC communication bandwidth simultaneously.

Let us first briefly describe how Gazelle implemented GC and point out the overhead.
The cloud computes linear operators using HE, then it adds random integers sampled from $\mathbb{Z}_t$ to the results, which are then sent to the proxy,
to prevent the proxy from learning any information on the true results, say $x$, after decryption (i.e. additive secret sharing). 
Now, the cloud and proxy have additive secret shares $(s_x, p_x)$ 
such that their summation is the true result of the linear function ($x = s_x+p_x \mod t$).
Note that, after cloud computation, since the true value $x$ is encrypted, the cloud does not know $x$,
but it can still add a random number $r$ to $x$ and generate secret shares $s_x = -r \mod t,\, p_x = x+r \mod t$.
Once the proxy receives its share $p_x$, 
the nonlinear activation function is computed jointly using GC, 
which is composed of addition of secret shares (to recover $x$ securely), 
nonlinear functions (e.g. ReLU or MaxPool), 
and addition of another random number $s_y$ (to hide the activation output).
However, this implementation has two overheads:
1) the precision of GC (i.e. activation precision) is fixed to HE parameter $t$ ($>$ 19 bits in Gazelle), and that much precision for all the layers is typically redundant in many fixed-point ML models~\cite{lin2016fixed}, and 
2) modular addition of secret shares incurs additional circuit overhead needed to address a wrap-around caused by modular operation (see Fig.~\ref{fig:opt_gc}). 

In light of these inefficiencies, as shown in Fig.~\ref{fig:opt_gc},
we propose truncating ($f$ LSBs in Fig.~\ref{fig:opt_gc}) in the secret shares instead of performing truncation using HE.
Then, the GC is built for the reduced number of bits 
to minimize its size, thereby saving communication bandwidth.
Note that this technique makes the protocol support fixed-point arithmetic without any overhead.
The required precision for activation in the model decides the number of bits used in GC.
In other words, as the model uses lower-precision activations, more bandwidth saving can be achieved.
Moreover, by ensuring that the maximum value of linear layer output is much smaller than $t$, 
we can further save the circuits (MUXs and subtractors in Fig.~\ref{fig:opt_gc}) originally needed to take care of wrap-around in modular operation.

\subsection{Overall Protocol} 

Below we describe the overall privacy-preserving CNN inference protocol,
Impala, which consists of five major parts split across the client, proxy, and cloud service provider.

\textbf{Key Generation (Client):}
The client generates a secret key for encryption and public keys required for automorphism and key switching in the cloud.
Note that this step is performed just once at the initial setup or when the client wants to generate new keys.

\textbf{Batch Encoding and Encryption (Client):}
The client packs/encrypts the private data and transmits it to the cloud.

\textbf{Computations on Encrypted Data (Cloud):}
After receiving encrypted data from the client,
the cloud performs Conv/FC computation to process
linear functions in the encrypted domain (see Section~\ref{SC}). 
For plaintext multiplication, weights in Conv and FC layers need to be packed into plaintexts.
Weight packing is input independent and can be done in advance to reduce online latency.
After computing linear function homomorphically, the cloud converts the result ciphertext
using the procedure described in Section~\ref{ks}
so the proxy can decrypt it with the proxy key.
Note that re-encryption is required only once (in the first layer)
regardless of the model depth since all subsequent layers are processed between only the proxy and the cloud using the proxy key. 
Finally, the cloud adds random numbers to the encrypted result before sending it to the proxy.

\textbf{Nonlinear Activation Function (Proxy):}
The proxy, using the proxy key, first decrypts the ciphertext received from the cloud
and obtains a plaintext polynomial, which needs to be unpacked before evaluating the nonlinear functions.
Unpacking returns a vector of $n$ secret shares, 
and the proxy and cloud jointly evaluates the activation output using GC.
After evaluating the nonlinear functions as explained in Section~\ref{gc},
the proxy packs the results and encrypts them using the proxy key.
The ciphertext is sent back to the cloud for evaluating the next layer in the network.
The cloud first first needs to subtract the random numbers ($s_y$ in Fig.~\ref{fig:opt_gc}) added to hide the activation outputs in GC and to recover the true outputs.
This process is repeated for each layer.
The sigmoid activation in the last layer is not necessary for inference
because sigmoid increases monotonically and the inference result can be obtained from the index of the maximum value of the last FC layer output~\cite{gilad2016cryptonets}.

\textbf{Data Decryption and Decoding (Client):}
Once evaluating the entire network is done,
the encrypted result is sent back to the client, who gets the inference result after decrypting and unpacking it.

\section{Experiments}
\label{sec:experiments}

\begin{table}[t]
\caption{Matrix-vector multiplication  comparison.}
\label{t:com_fc}
\begin{center}
\begin{small}
\begin{sc}
\begin{tabular}{ccccc}
\toprule
 & \multicolumn{2}{c}{All Keys} & \multicolumn{2}{c}{$\log n$ Keys} \\
Key Size & 629MB & 189MB & 3.4MB & 1.0MB \\
\midrule
Matrix Size & Gazelle & Impala & Gazelle & Impala \\

\midrule
256 $\times$ 128 & 0.017  & 0.005  & 0.038  &  0.007 \\
1024 $\times$ 128 & 0.052 & 0.017 & 0.19 & 0.034  \\
4096 $\times$ 1024 & 2.23 & 0.51 & 13.4 & 1.62  \\
\bottomrule
\end{tabular}
\end{sc}
\end{small}
\end{center}
\end{table}

\begin{table}[t]
\caption{Convolution comparison.}
\label{t:com_conv}
\begin{center}
\begin{small}
\begin{sc}
\begin{tabular}{ccccc}
\toprule
 & \multicolumn{2}{c}{All Keys} & \multicolumn{2}{c}{$\log n$ Keys} \\
Key Size & 629MB & 189MB & 3.4MB & 1.0MB \\
\midrule
$(w,c_i,f_w,c_o)$ & Gazelle & Impala & Gazelle & Impala \\
\midrule
$(32,32,3,32)$ & 2.04 & 0.92 & 5.14 & 1.14 \\
$(16,128,3,128)$ & 8.5 & 3.68 & 20.6 & 4.57 \\
$(64,64,3,64)$ & 173 & 38 & 316 & 45.8 \\
\bottomrule
\end{tabular}
\end{sc}
\end{small}
\end{center}
\end{table}

\begin{table}[t]
\caption{Activation comparison (for 10,000 outputs).}
\label{t:com_act}
\begin{center}
\begin{small}
\begin{sc}
\begin{tabular}{ccccc}
\toprule
  & \multicolumn{2}{c}{Gazelle} & \multicolumn{2}{c}{Impala} \\
  & Offline & Online & Offline & Online \\
\midrule
ReLU & 0.57s & 0.13s & 0.45s & 0.05s \\
  & 49.3MB & 15.2MB & 12.8MB & 8MB \\
MaxPool & 1.37s & 0.36s & 0.94s & 0.16s \\
  & 141.5MB & 51.7MB & 41.6MB & 27.2MB \\
\bottomrule
\end{tabular}
\end{sc}
\end{small}
\end{center}
\end{table}

\begin{table*}[t]
\caption{Client per-inference communication overhead comparison with state-of-the-art.}
\label{t:comp_overhead}
\begin{center}
\begin{small}
\begin{sc}
\begin{tabular}{cccccc}
\toprule
Communication BW [MB] & CryptoNets & LoLa & MiniONN & Gazelle & Impala \\
\midrule
MNIST & 372 & 5.4 (est.)  & 637 & 70 & {\bf 0.061 ($>$ 80 $\times$)} \\
CIFAR-10 & N/A & 340 (est.) & 6226 & 1236 & {\bf 0.092 ($>$ 3500 $\times$)} \\
\bottomrule
\end{tabular}
\end{sc}
\end{small}
\end{center}
\end{table*}
\begin{table*}[t]
\caption{Overall per-inference performance summary and comparison.}
\label{t:mnist}
\begin{center}
\begin{small}
\begin{sc}
\begin{tabular}{ccccccc}
\toprule
MNIST & Runtime [s] & Communication [MB] & Accuracy [\%] \\
\midrule
CryptoNets~\cite{gilad2016cryptonets}  & 298 & 372 & 98.95 \\
Faster CryptoNets~\cite{chou2018faster}       & 39.1 & 411 & 98.71 \\
LoLa~\cite{brutzkus2019low} & 2.2 & 5.4 (est.) & 98.95 \\
MiniONN~\cite{liu2017oblivious} & 5.74 & 637 & 99.31  \\
Gazelle~\cite{juvekar2018gazelle} & 0.81  & 70 & N/A \\
Impala & {\bf 0.27} & {\bf 23} & {\bf 99.22} \\
\bottomrule
\toprule
CIFAR-10 & Runtime [s] & Communication [MB] & Accuracy [\%] \\
\midrule
LoLa~\cite{brutzkus2019low}  & 720 & 340 (est.) & N/A \\
MiniONN~\cite{liu2017oblivious} & 72 & 6226 & 81.61  \\
Gazelle~\cite{juvekar2018gazelle} & 12.9  & 1236 & N/A \\
Impala & {\bf 3.3} & {\bf 364} & {\bf 81.63} \\
\bottomrule
\end{tabular}
\end{sc}
\end{small}
\end{center}
\vspace{-1em} 
\end{table*}

The HE encryption parameters for the experiments are chosen to meet 128-bit security parameter, same as those in \cite{juvekar2018gazelle} with 19-bit plaintext and 60-bit ciphertext modulus. 
We implemented Impala in C++ with SEAL library~\cite{sealcrypto}, which contains basic functions used in the BFV scheme, and Multi-Protocol SPDZ~\cite{mpspdz} for the GC implementation. To evaluate the performance, we ran the models on Intel Xeon Gold 5215 2.50\,GHz CPUs, and the communication links between the parties are in the LAN setting similar to previous works.

In order to show the impact of each technique proposed in Section~\ref{sec:proposed},
performance comparison of each operator is first presented.
Table~\ref{t:com_fc} and Table~\ref{t:com_conv} show the latency of Gazelle and Impala
when they compute matrix-vector multiplication and convolution, respectively.
For both, we also compare the latency between two cases: 1) all the automorphism keys (left columns) and 2) only $\log n$ keys (right columns) are provided. 
In the case when all the keys are provided, Gazelle can take advantage of the hoisting optimization and improves latency by reducing the required number of NTTs,
but this imposes a large burden on the client to generate keys with size more than 600$\times$ than Impala.
As shown in Table~\ref{t:com_fc} and Table~\ref{t:com_conv}, Gazelle's latency degrades significantly without the client-side burden.
In both cases, Impala shows the speed-up up to 8.3$\times$ and 7$\times$ over Gazelle.

Table~\ref{t:com_act} shows the performance comparison between Gazelle and Impala for evaluating 10,000 outputs of ReLU and $(2\times2)$-MaxPool,
widely-used activations in state-of-the-art CNN models.
In this comparison, GC of Gazelle is built for 19-bit input secret shares (same as plaintext modulus),
while GC of Impala is built for 10-bit inputs after truncating 9 LSBs of secret shares,
which are the parameters we used for evaluating overall CNN models shown below.
Due to the smaller GC size, the offline and online communication bandwidth is reduced to around 1/4 and 1/2, respectively.

To compare the overall CNN inference, we evaluated the performance of two CNN models of different scale 
using the MNIST and CIFAR-10 datasets. 
For MNIST we use a 4-layer CNN with 2-Conv and 2-FC,
and our CIFAR-10 model is an 8-layer CNN with 7-Conv and 1-FC.
For fair comparison with other state-of-the-art, we use the same CNN architectures as Gazelle. 

Table~\ref{t:comp_overhead} shows how much client communication overhead can be reduced 
using Impala compared to other state-of-the-art for both models. 
As shown in Table~\ref{t:comp_overhead}, by introducing the proxy securely in the protocol using {\em re-encryption with key switching},
it significantly reduces the amount of communication bandwidth on the client side for a single inference task. 
Note that, with Impala, 
the client overhead is only dependent on the data size
and is irrelevant of the model architecture (e.g., kernel size, model depth)
since the client is not involved during the computation process. 
Although MNIST and CIFAR-10 models are evaluated in this paper, 
deeper models for ImageNet such as ResNet50~\cite{he2016deep} will only affect the runtime and bandwidth of the cloud and the proxy, 
not those of the client, providing a scalable solution for mobile cloud computing services.
Thanks to the proposed low-overhead re-encryption, the client can save more than orders of magnitude, 
while adding a negligible latency overhead of less than 5\,ms in the cloud.

Both overall inference runtime (latency) and the total communication bandwidth required for a single inference are compared in Table~\ref{t:mnist}. 
Note that the numbers shown in Table~\ref{t:mnist} correspond to single inference with batch size of 1, i.e., single image inference.
To show the problems of different solutions,
recent HE-based (CryptoNets and LoLa), MPC-based (MiniONN), and (HE + GC)-based (Gazelle) protocols 
are picked as baseline.
The proposed sparse homomorphic convolution and GC implementation reduce 
both computational and communication overheads incurred by HE and GC respectively,
resulting in significant performance improvement.
As a result, Impala provides overall bandwidth savings of more than $3\times$ (more than 4 orders of magnitude saving in the client) and speedup of $4\times$, compared to the current state-of-the-art techniques.

\section{Conclusion}
\label{conclusion}

This paper proposes Impala, a low-latency communication-efficient protocol for privacy-preserving deep learning inference.
Several techniques tailored for deep learning inference ---{\em re-encryption with key switching}, {\em sparse homomorphic convolution}, and privacy-preserving activation evaluation using {\em garbled circuit with truncated secret shares}---to reduce computational and communication overhead incurred by cryptography were proposed. 
Impala provides bandwidth savings of more than $3\times$ and speedup of $4\times$
while reducing client overhead by several orders of magnitude, 
compared to the current state-of-the-art technique.



\bibliography{refs}
\bibliographystyle{mlsys2020}

\ignore{
\appendix
\section{Please add supplemental material as appendix here}
Put anything that you might normally include after the references as an appendix here, {\it not in a separate supplementary file}. Upload your final camera-ready as a single pdf, including all appendices.
}

\end{document}